\documentclass[aps,prl,twocolumn,showpacs,amsmath,amssymb,superscriptaddress,letterpaper]{revtex4}
\usepackage{times}
\usepackage{amsfonts}
\usepackage{mathrsfs}
\usepackage{graphicx}
\usepackage{dcolumn}
\usepackage{bm}
\usepackage{color}

\usepackage[colorlinks,bookmarks=false,citecolor=blue,linkcolor=red,urlcolor=blue]{hyperref}
\def\be{\begin{equation}}       \def\ee{\end{equation}}
\def\bea{\begin{eqnarray}}      \def\eea{\end{eqnarray}}

\begin{document}
\title{$g$-wave  Pairing  in BiS$_2$ Superconductors }

\author{Xianxin Wu}
\affiliation{ Institute of Physics, Chinese Academy of Sciences,
Beijing 100190, China}

\author{Jing Yuan}
\affiliation{ Institute of Physics, Chinese Academy of Sciences,
Beijing 100190, China}

\author{Yi Liang}
\affiliation{ Institute of Physics, Chinese Academy of Sciences,
Beijing 100190, China}

\author{Heng Fan }  \affiliation{ Institute
of Physics, Chinese Academy of Sciences, Beijing 100190, China}

\author{Jiangping Hu  }\email{jphu@iphy.ac.cn} \affiliation{
Institute of Physics, Chinese Academy of Sciences, Beijing 100190,
China}\affiliation{Department of Physics, Purdue University, West
Lafayette, Indiana 47907, USA}

\date{\today}

\begin{abstract}
Recent angle resolved photoemission spectroscopy(ARPES) experiments have suggested that  BiS$_2$ based superconductors are at very low electron doping.   Using random phase approximation(RPA) and functional renormalization group(FRG) methods, we find that $g$-wave pairing symmetry belonging to A$_{2g}$ irreducible representation is  dominant at electron doping $x<0.25$.  The pairing symmetry  is determined by inter-pocket nesting and orbital characters on the Fermi surfaces and is robust  in a two-orbital model including both Hund's coupling $J$, and Hubbard-like Coulomb interactions $U$ and $U'$ with relatively small $J$ ($J\leq0.2U$).    With the increasing electron doping,  the g-wave state competes with both  the s-wave $A_{1g}$  and d-wave $B_{2g}$ states and no pairing symmetry emerges dominantly.
\end{abstract}


\maketitle
Recently, a new family of materials containing BiS$_2$ layers
has been discovered to be superconducting (SC) and drawn many research attentions\cite{Mizuguchi2012,Singh2012,Mizuguchi2,Demura,Xing,Jha,Lin2013,Yazici2013,Deguchi2013,Biswas2013,YKLi2014}.
Similar to cuprate and iron based superconductors, these materials consist of
two dimensional BiS$_2$ layers and various  types of blocking layers.  The essential electronic properties  are attributed to the BiS$_2$ layers.
According to density functional theory calculations, the parent compounds of the BiS$_2$-based superconductors are semiconducting and  the conduction
band is mainly attributed to the 6$p$
orbitals of Bi\cite{Mizuguchi2012,Usui2012}. A two-orbital model, including Bi $p_X$ and $p_Y$ orbitals, reasonably   describes the band structure\cite{Usui2012} that controls major electronic properties.  Superconductivity is induced by electron doping.

Due to the weak correlation effect in $p$ orbitals and the low superconducting transition temperature, electron phonon
coupling has been suggested to play a dominant role in superconducting pairing\cite{Wan2013,Li2013,Yildirim2013}. However, in the recent neutron scattering experiment, the observed almost unchanged low-energy modes indicated that the electron phonon coupling could be much weaker than expected\cite{Lee2013}. Moreover, the large ratio $2\Delta/T_c$ may suggest that the pairing mechanism is unconventional\cite{SLi2013,Liu2013}. Electron-electron correlations can be responsible for the Copper pairing\cite{Zhou2013,Liang2014,Matrins2013,Yang2013}.

When electron-electron correlations are the driving force for pairing, unconventional pairing symmetries arise naturally.  There have been quite a few theoretical studies about possible pairing symmetries in these new superconductors\cite{Zhou2013,Liang2014,Matrins2013,Yang2013}.  As the nominal compositions of the  superconducting  materials  indicated high electron doping\cite{Mizuguchi2,YKLi2014}, all these previous studies concentrated on the high electron doping region  where the  electronic structure was featured with large Fermi surfaces(FS) at $\Gamma$ and $M$ points in Broullioun zone (BZ) in close vicinity to Van Hove singularity. However,   very recently, two ARPES groups have reported that  there are only two small electron pockets around X points \cite{Zeng2014,Ye2014} and the true electron fillings are much smaller than those  expected from the nominal compositions.

In this paper, we investigate the pairing symmetry of BiS$_2$ based superconductors at low doping level using random phase approximation(RPA) and functional renormalization group(FRG) methods. We find that $g$-wave  pairing state that belongs to the A$_{2g}$ irreducible representations  of the lattice symmetry is dominant at electron doping $x<0.25$ in a two-orbital model including both Hund's coupling $J$, and Hubbard-like Coulomb interactions $U$ and $U'$  in the reasonable parameter region $J\leq0.2U$. This robust pairing symmetry is determined by inter pocket nesting and orbital characters on the FS. With the increase of  $J$ or electron doping level,  the g-wave loses its dominance and competes with other pairing symmetries.  In both cases, there is no single dominant pairing wave.  For example, with a large $J$, a d-wave ($B_{1g}$) is only slightly favored over  a s-wave ($A_{1g}$)  and   the g-wave, and at high doping near the  Lifshitz transition on which  the previous studies concentrated,  the s-wave   and the other d-wave ($B_{2g}$) are almost equally favored. Due to the close competition between s-wave and d-wave, we speculate that superconductivity may not takes place at high electron doping. Our results, therefore, predict a new pairing symmetry for the BiS$_2$ superconductors. As the g-wave pairing state has a distinct nodal structure on FS, our prediction can  be experimentally tested.

\begin{figure}
 \includegraphics[height=8.0cm]{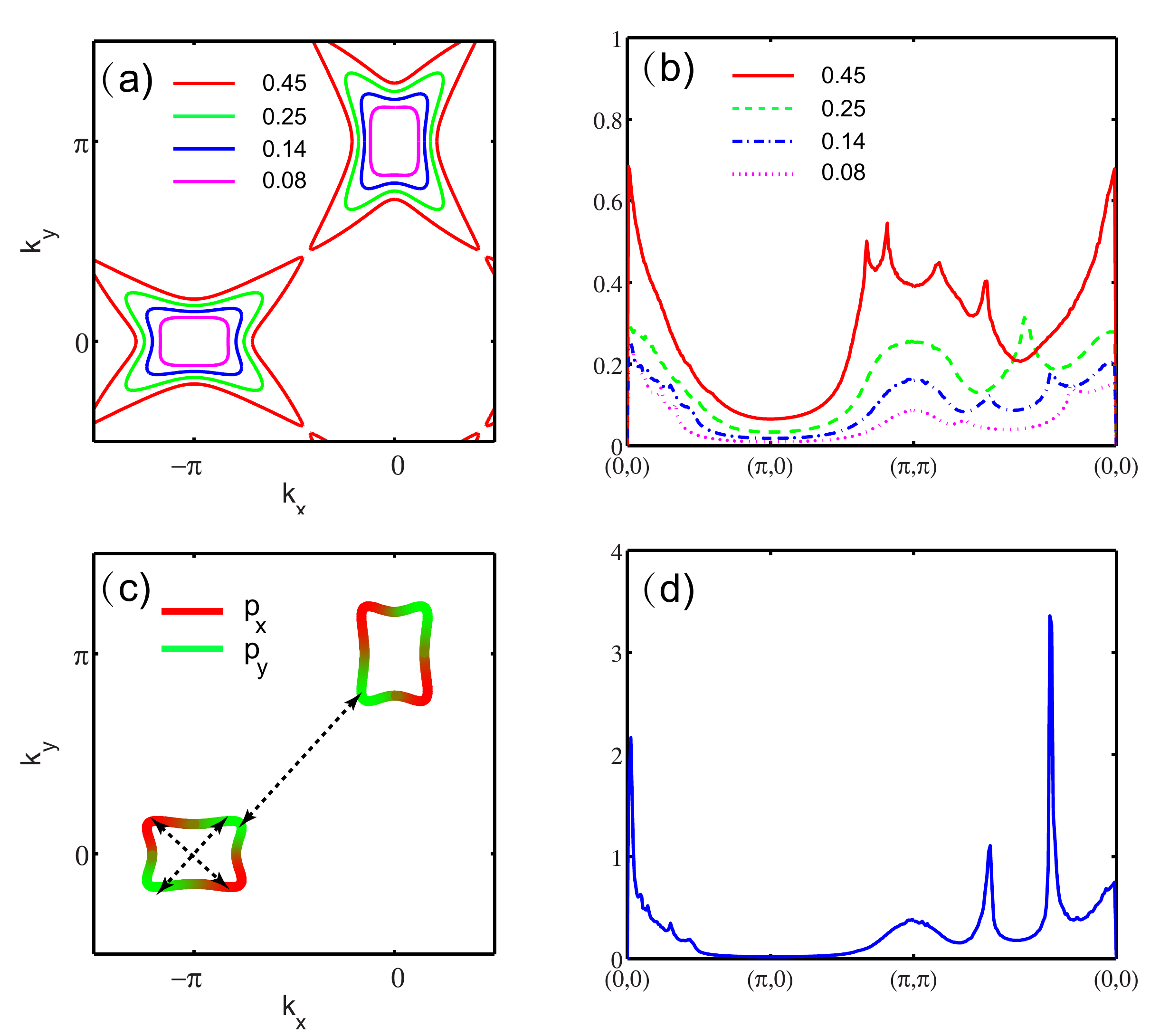}
 \caption{ (Color online) (a) Fermi surfaces for different electron doping. (b) The bare susceptibility for the same electron fillings as in (a). (c) The orbital characters on FS for $x$=0.14. The main contributions are shown by the following colors: red for $p_X$ and green for $p_Y$. The FS nesting vectors are indicated by black lines with arrows. (d) The RPA susceptibility for $x$=0.14 with U=3.5 and J=0. The positions of two sharp peaks on the right correspond to the nesting vectors shown in (c). }\label{FS_sus}
 \end{figure}

We adopt the two-band model, the tight binding Hamiltonian\cite{Usui2012} is $H_{0}=\sum_{\mathbf{k}\sigma}\Psi_{\mathbf{k}\sigma}^{\dagger}T(\mathbf{k})\Psi_{\mathbf{k}\sigma}$,
\begin{equation}
T(\mathbf{k})=\left(\begin{array}{cc}
\epsilon_{X}(\mathbf{k})-\mu & \epsilon_{XY}(\mathbf{k})\\
\epsilon_{XY}(\mathbf{k})^{*} & \epsilon_{Y}(\mathbf{k})-\mu\end{array}\right)\end{equation}
where $\Psi_{\mathbf{k}\sigma}^{\dagger}=(c_{X\mathbf{k}\sigma}^{\dagger},c_{Y\mathbf{k}\sigma}^{\dagger})$
is the creation operator for spin $\sigma$ electrons in the two orbitals $p_{X},p_{Y}$ and $\epsilon_X(\textbf{k})$, $\epsilon_Y(\textbf{k})$ and $\epsilon_{XY}(\textbf{k})$ are the same as those defined in Ref.\cite{Liang2014}. As the observed electron doping is much less than those inferred from the nominal composition, we discuss the pairing properties based on  FS before the Lifshitz transition. For the case after Lifshitz transition, RPA calculations have been done in Ref.\cite{Matrins2013}. Fig.\ref{FS_sus}(a) shows the FS electron pockets for four different electron fillings $x=$0.08, 0.14, 0.25 and 0.45. The corresponding bare spin susceptibilities are shown in Fig.\ref{FS_sus}(b). There is a rectangle-shaped electron pocket centered at each X point for $x=0.14$. Compared with experimental data\cite{Zeng2014,Ye2014}, a smaller electron pocket at X point is absent because the real materials contain two BiS$_2$ layers but the model is based on a single BiS$_2$ layer and neglects the interlayer coupling.  The orbital characters on FS are shown in Fig.\ref{FS_sus}(c), where the colors correspond to the dominant orbital weight(red for $p_X$ and green for $p_Y$). The two right peaks (in blue dash dot line) at $\textbf{q}_1=(0.62\pi,0.62\pi)$ and $\textbf{q}_2=(0.32\pi,0.32\pi)$ correspond to inter and intra FS nesting, respectively. The broad peak at ($\pi$,$\pi$), interpreted as inter pocket nesting, resembles the FS nesting in iron based superconductors\cite{Graser2009}. The nesting wave vectors are shown in Fig.\ref{FS_sus}(c). When interactions are introduced, the peaks at $\textbf{q}_1$ and $\textbf{q}_2$ in $\chi_0$ are the ones that diverge in the RPA spin susceptibility(Fig.\ref{FS_sus}(d)). With the increasing of electron doping, $\textbf{q}_1$ moves right but $\textbf{q}_2$ moves left. Near the Lifshitz transition point($x$=0.45), the broad peak disappears and many peaks appear in $\chi_0$ due to the inter pocket FS nesting. The RPA spin susceptibility diverges at certain $\textbf{k}$, indicating that the system is  unstable  to a long range magnetic order.
At low doping concentration ($x<0.2$), the FS are rectangle-shaped. The orbital characters on FS and inter pockets nesting clearly play a crucial role in determining the pairing symmetry.

The interaction part of Hamiltonian for this  multi-orbital system is,
\begin{eqnarray}
H_{int}&=&U\sum_{i\alpha}n_{i\alpha\uparrow}n_{i\alpha\downarrow}+U'\sum_{i,\alpha<\beta}n_{i\alpha}n_{i\beta} \nonumber\\
&+&J\sum_{i,\alpha<\beta,\sigma\sigma'}c^{\dag}_{i\alpha\sigma}c^{\dag}_{i\beta\sigma'}c_{i\alpha\sigma'}c_{i\beta\sigma}
+J'\sum_{i,\alpha\neq\beta}c^{\dag}_{i\alpha\uparrow}c^{\dag}_{i\alpha\downarrow}c_{i\beta\downarrow}c_{i\beta\uparrow}
\end{eqnarray}
where $n_{i\alpha}=n_{\alpha\uparrow}+n_{\alpha\downarrow}$. Here we adopt  the   parameter notations given in Ref.\cite{Kemper2010}. The effective interaction obtained in the RPA approximation is,
\begin{eqnarray} V_{eff}=\sum_{ij,\textbf{k}\textbf{k}'}\Gamma_{ij}(\textbf{k},\textbf{k}')c^{\dag}_{i\textbf{k}\uparrow}c^{\dag}_{i-\textbf{k}\downarrow}c_{j-\textbf{k}'\downarrow}c_{j\textbf{k}'\uparrow}
\end{eqnarray}
where the momenta $\textbf{k}$ and $\textbf{k}'$ are restricted to  different FS $C_i$ with $\textbf{k}\in C_i$ and $\textbf{k}'\in C_j$ and $\Gamma_{ij}(\textbf{k},\textbf{k}')$ is the pairing scattering vertex in the singlet channel\cite{Kemper2010}. The pairing vertex is,
\begin{eqnarray}
\Gamma_{ij}(\textbf{k},\textbf{k}')=Re[\sum_{l_1 l_2 l_3 l4}a^{l_2,*}_{v_i}(\textbf{k}) a^{l_3,*}_{v_i}(-\textbf{k}) \times \nonumber\\
 \Gamma_{l_1 l_2 l_3 l_4}(\textbf{k},\textbf{k}',\omega=0) a^{l_1}_{v_j}(\textbf{k}') a^{l_4}_{v_j}(-\textbf{k}')],
\end{eqnarray}
 where $a^{l}_{v}$(orbital index $l$ and band index $v$) is the component of the eigenvectors from the diagonalization of the tight binding Hamiltonian. The orbital vertex function $\Gamma_{l_1 l_2 l_3 l_4}$ in the fluctuation exchange formulation\cite{Bickers1989,Kubo2007,Kemper2010} are given by,
 \begin{eqnarray}
\Gamma_{l_1 l_2 l_3 l_4}(\textbf{k},\textbf{k}',\omega)=[\frac{3}{2}\bar{U}^s \chi^{RPA}_1(\textbf{k}-\textbf{k}',\omega)\bar{U}^s+\frac{1}{2}\bar{U}^s \nonumber\\
-\frac{1}{2}\bar{U}^c\chi^{RPA}_0(\textbf{k}-\textbf{k}',\omega)\bar{U}^c+\frac{1}{2}\bar{U}^c]_{l_3 l_4 l_1 l_2}.
\end{eqnarray}
The $\bar{U}_s$ is the spin interaction matrice and the $\bar{U}_c$ the charge spin interaction matrice, defined in Ref.\cite{Kemper2010}. The $\chi^{RPA}_0$ describes the charge fluctuation contribution and the $\chi^{RPA}_1$ the spin fluctuation contribution. For a given gap function $g(\textbf{k})$, the pairing strength functional is,
\begin{eqnarray}
\lambda[g(\textbf{k})]=-\frac{\sum_{ij}\oint_{C_i} \frac{dk_{\|}}{v_F(\textbf{k})} \oint_{C_j} \frac{dk'_{\|}}{v_F(\textbf{k}')} g(\textbf{k})\Gamma_{ij}(\textbf{k},\textbf{k}') g(\textbf{k}')} {4\pi^2\sum_i\oint_{C_i} \frac{dk_{\|}}{v_F(\textbf{k})} [g(\textbf{k})]^2 },
\label{strength}
\end{eqnarray}
where $v_F(\textbf{k})=|\bigtriangledown_{\textbf{k}}E_i(\textbf{k})|$ is the Fermi velocity on a given fermi surface sheet $C_i$. $g(\textbf{k})$ is determined as the station solution of Eq.\ref{strength}. The obtained gap function should have the symmetry of one of the irreducible representations of the corresponding point group. Although the point group for the BiS$_2$ layer is $C_{4v}$, the point group symmetry in our effective model is $D_{4h}$.  We consider   one dimensional irreducible representations $A_{1g}$, $A_{2g}$, $B_{1g}$ and $B_{2g}$. We perform calculations in the spin-rotational invariance case, where $U'=U-2J$ and $J=J'$. The typical temperature $T=0.02$  is used and $\eta=0.005$ is adopted to regularize the Green's functions. All the summations over the Brillouin Zone are performed with uniform $200\times200$ meshes.

\begin{figure}
 \includegraphics[height=12cm]{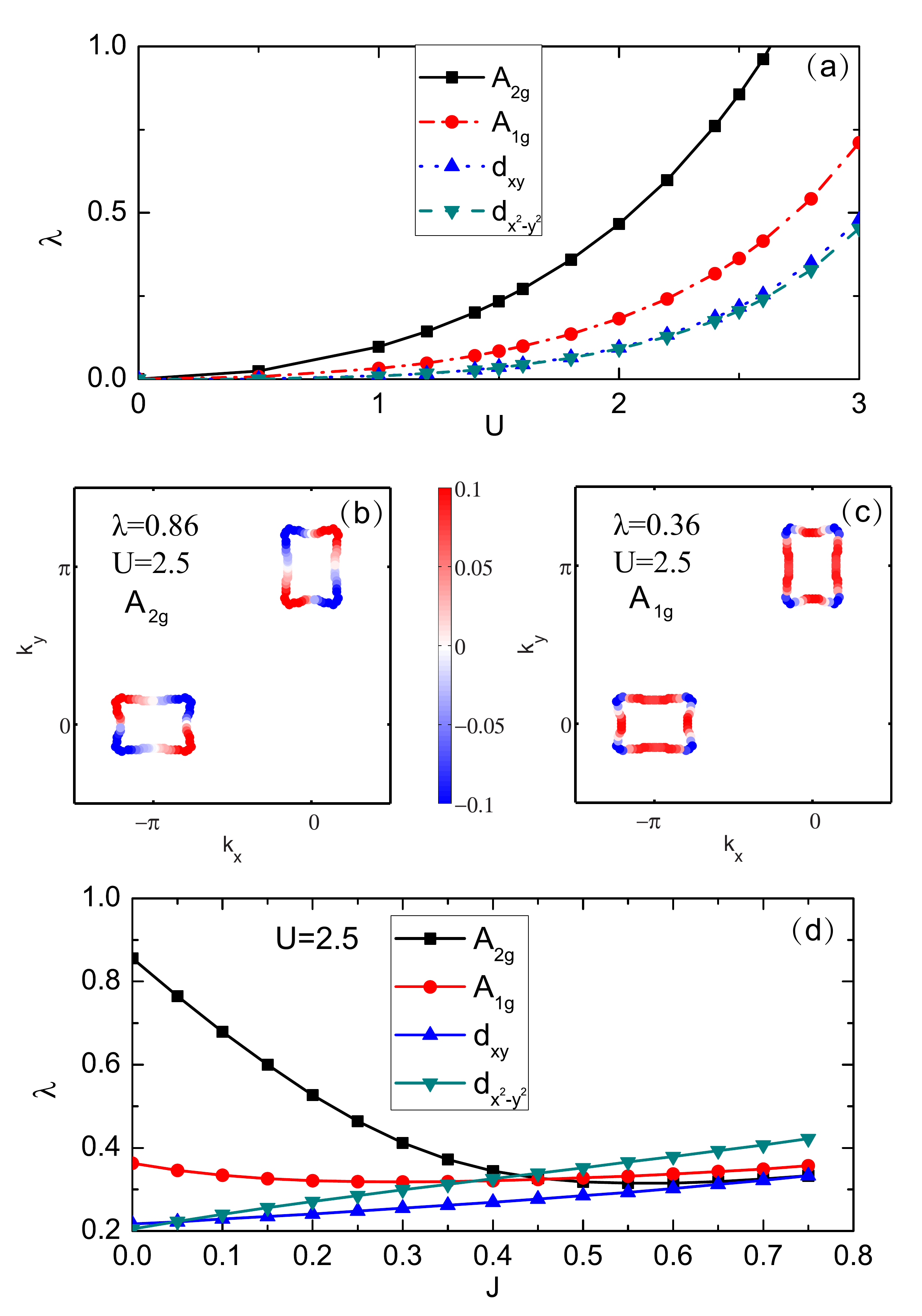}
 \caption{ (Color online) The pairing strengths $\lambda$ and gap functions for the electron doped compound($x$=0.14) for $J=0$. (a) The four largest pairing strengths as a function of $U$. The two dominant gap functions for $A_{2g}$ (b) and $A_{1g}$ (c), calculated close to the instability (U=2.5). (d) The four largest pairing strengths as a function of $J$ with U=2.5. }\label{UJ0}
 \end{figure}

First, we consider the case where the Hund's rule coupling is negligible compared with the intraorbital Coulomb interaction. Fig.\ref{UJ0}(a) shows the pairing strength eigenvalues for the four leading eigenvalues as a function of $U$ at  the electron doping $x$=0.14. We find that the dominant gap function has the symmetry $A_{2g}$ and the order parameter of this  state is shown in Fig.\ref{UJ0}(b) for $U=2.5$. This pairing symmetry is $g$-wave, which changes sign 8 times in a $2\pi$ rotation. This state is odd over all the mirror reflections($x$,$y$ and diagonal reflections). Therefore, nodes appear in the $k_{x/y}=0$, $\pi$ lines. There is also a sign change within FS sheets. Fig.\ref{UJ0}(c) shows the subdominant gap function which has $A_{1g}$ symmetry. This extended $s$-wave state features a sign change and an anisotropic gap distribution on the FS sheets. We can calculate the contributions of intra-sheet and inter-sheet scattering processes for the two leading pairing states separately. We find that the intra-sheet process contributes negatively to the g-wave state while  inter-sheet process contributes positively. Both of them show rapid increase with the increasing of $U$. As the inter-sheet scattering always overcomes the intra-sheet scattering, the $A_{2g}$ is stable.  The contribution of both intra-sheet and inter-sheet processes are positive for the $A_{1g}$ s-wave.  The strong inter-pocket nesting results in a sign change of the superconducting order between the green $p_X$ (red $p_Y$) regions on the two electron pockets shown in Fig.\ref{FS_sus}(c). This is the essential reason why  the  g-wave symmetry is more stable than the s-wave with the  increasing of $U$.

\begin{figure}
 \includegraphics[height=4cm]{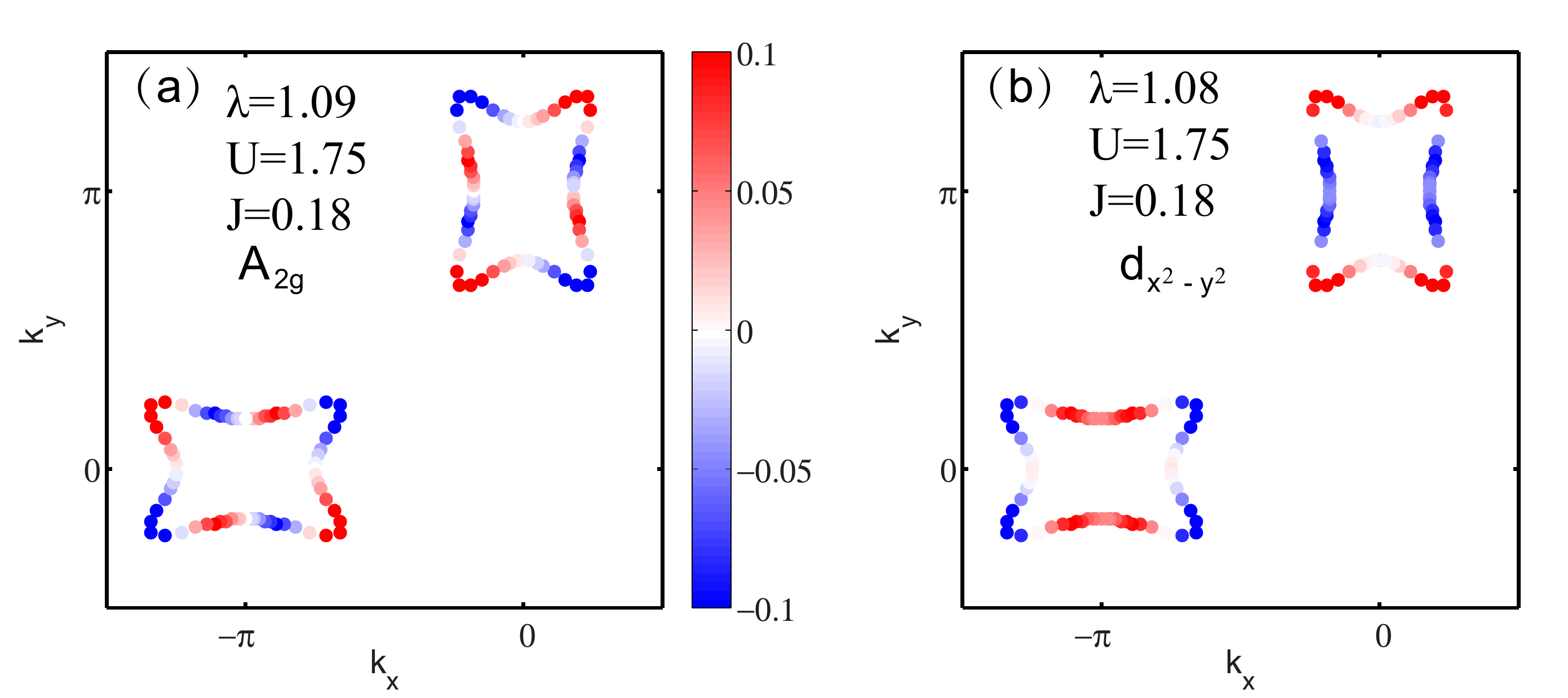}
 \caption{ (Color online) The two dominant gap functions with symmetries $A_{2g}$ (a) and $B_{1g}$ (b) at $x$=0.25, calculated with $U=1.75$ and $J=0.18$. The two pairing states are almost degenerate. }\label{UJ_0.25}
 \end{figure}

Second, we consider the effect of Hund's  coupling on the pairing symmetry. Fig.\ref{UJ0}(d) shows the pairing strength $\lambda$ for the four leading eigenvalues as a function of $J$ with $U=2.5$ and $x=0.14$.  The figure shows  that $J$ has a significant effect on the g-wave state but a negligible effect on the s-wave state. Around $J\sim 0.55$, the g-wave  pairing and  a  $B_{1g}$($d_{x^2-y^2}$) pairing become equally favored. The d-wave pairing state (not shown) is quite similar to that shown in Fig.\ref{UJ_0.25}(b). There is no sign change   on the same pocket  but a sign change between the two pockets. Due to the enhancement of inter orbital scattering with the increasing of $J$, both intra and inter orbital scattering in the inter-pocket processes become important. Then, the system favors a gap with a sign change between the FS sheets. Consequently, the gap with symmetry $d_{x^2-y^2}$ is favored.  However,  as shown in Fig.\ref{UJ0}(d), the pairing strength eigenvalues of the four leading state are very close to each other, indicating the intense competition between those states.

\begin{figure}
 \includegraphics[height=4cm]{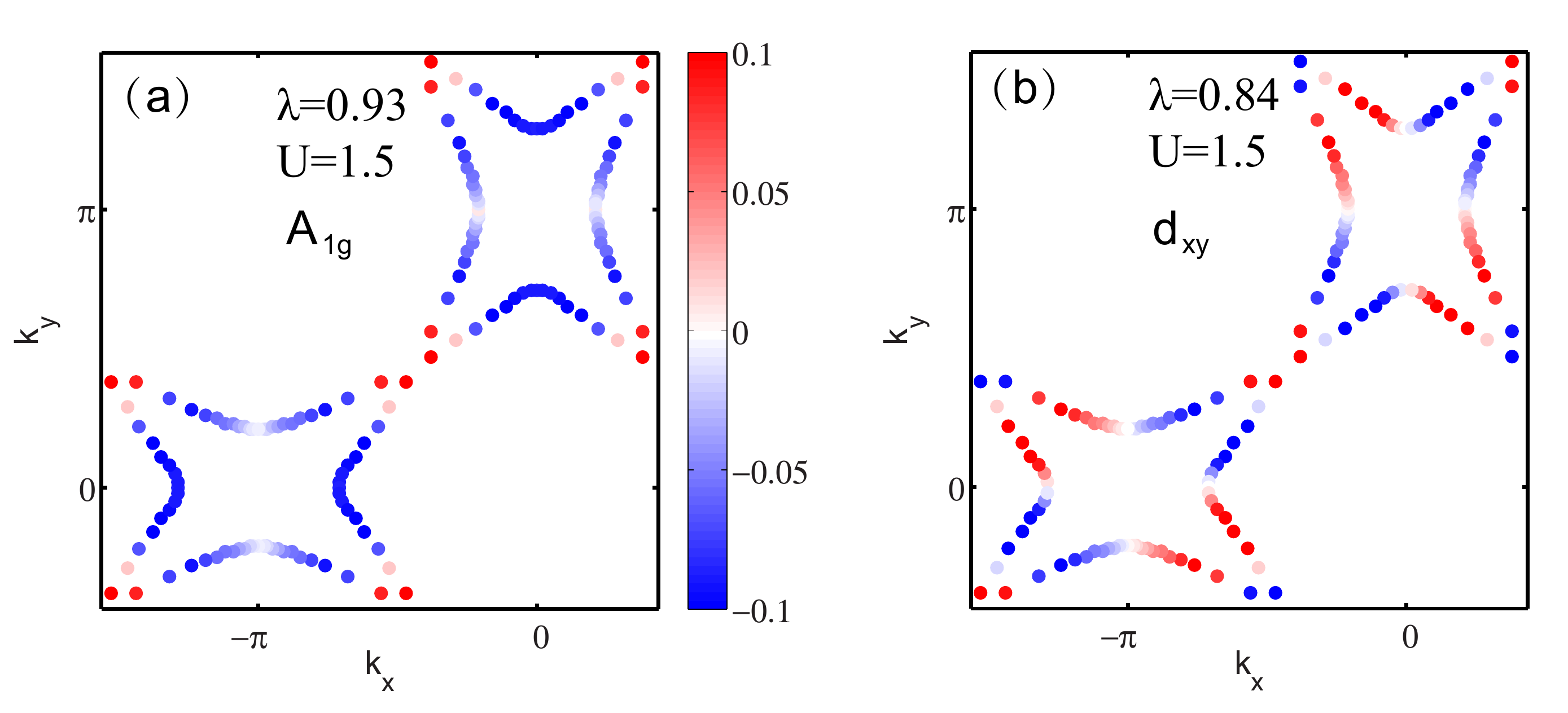}
 \caption{ (Color online) The two dominant gap functions with symmetries $A_{1g}$ (a) and $B_{2g}$ (b) at $x$=0.45, calculated with $U=1.5$ and $J=0$.}\label{UJ0_0.45}
 \end{figure}

When the electron filling is less than 0.14, we find the g-wave state is always strongly favored if $J$ is relative small. With the increasing of electron doping, the pockets enlarge and the shapes deviate from rectangle. The inter-pocket nesting becomes weaker while the intra-pocket nesting becomes stronger, which greatly affects the pairing strength of the g-wave pairing.  In fact,  we find that  the g-wave  and the $d_{x^2-y^2}$-wave ($B_{1g}$)  are almost equally favored at $x=0.25$ when $U=1.75$ and $J=0.17$. The two leading gap functions are shown in Fig.\ref{UJ_0.25}. Near the Lifshitz transition point($x$=0.45), the two leading states are  the s-wave $A_{1g}$ and  the d-wave $B_{2g}$, shown in Fig.\ref{UJ0_0.45}. The symmetries of the two leading gaps are the same as those at higher electron doping(after Lifshitz transition), studied  in Ref.\cite{Matrins2013}. Nevertheless, there are always strong competitions among multi-pairing channels at high electron doping. There is no obviously leading pairing symmetry. %

\begin{figure}
 \includegraphics[height=6cm]{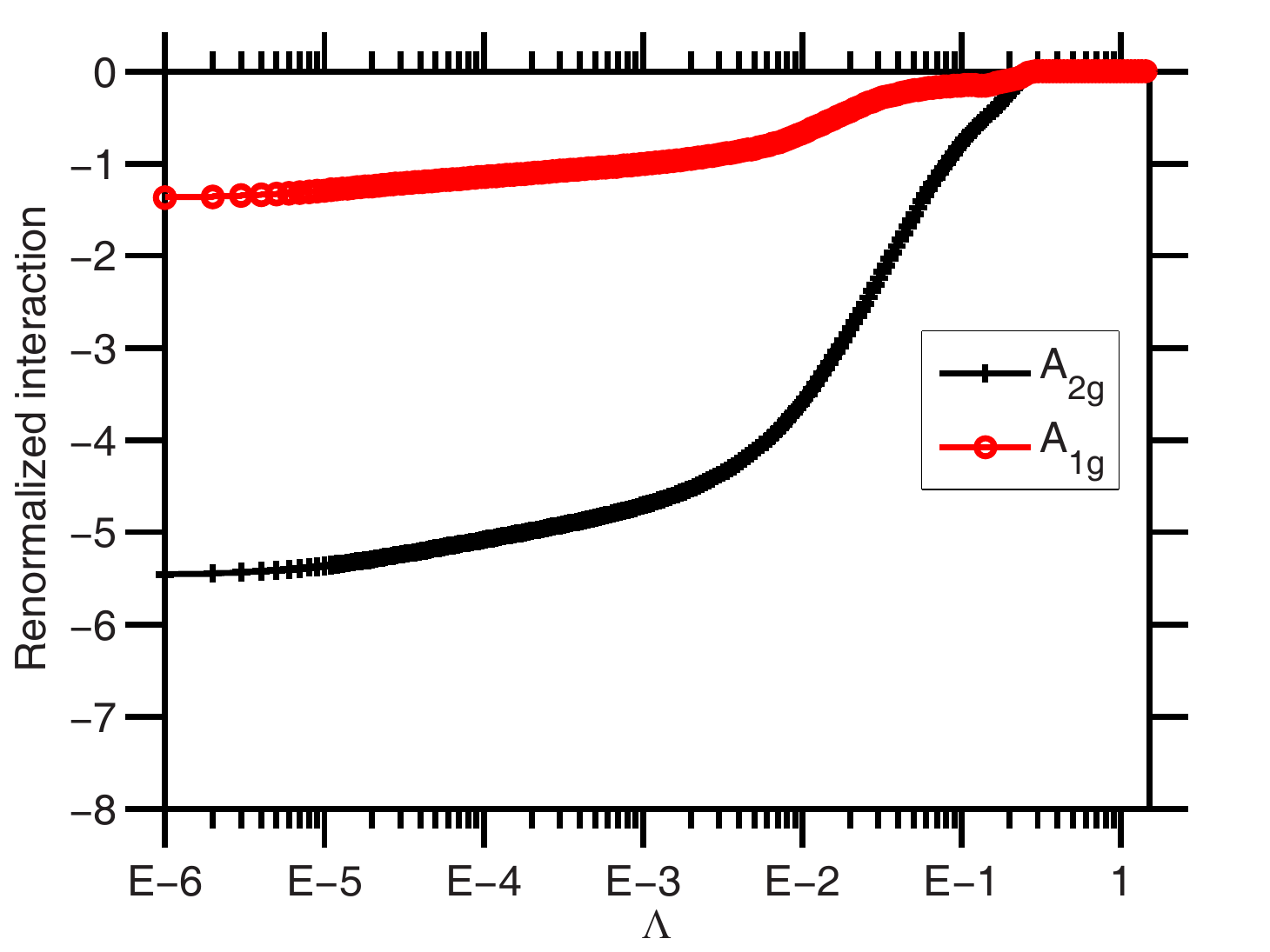}
 \caption{ (Color online) FRG flow of two leading pairing channels for the electron doped compound($x$=0.14) with U=2.5 and J=0.25. }\label{frg_flow}
 \end{figure}

\begin{figure}
 \includegraphics[height=4cm]{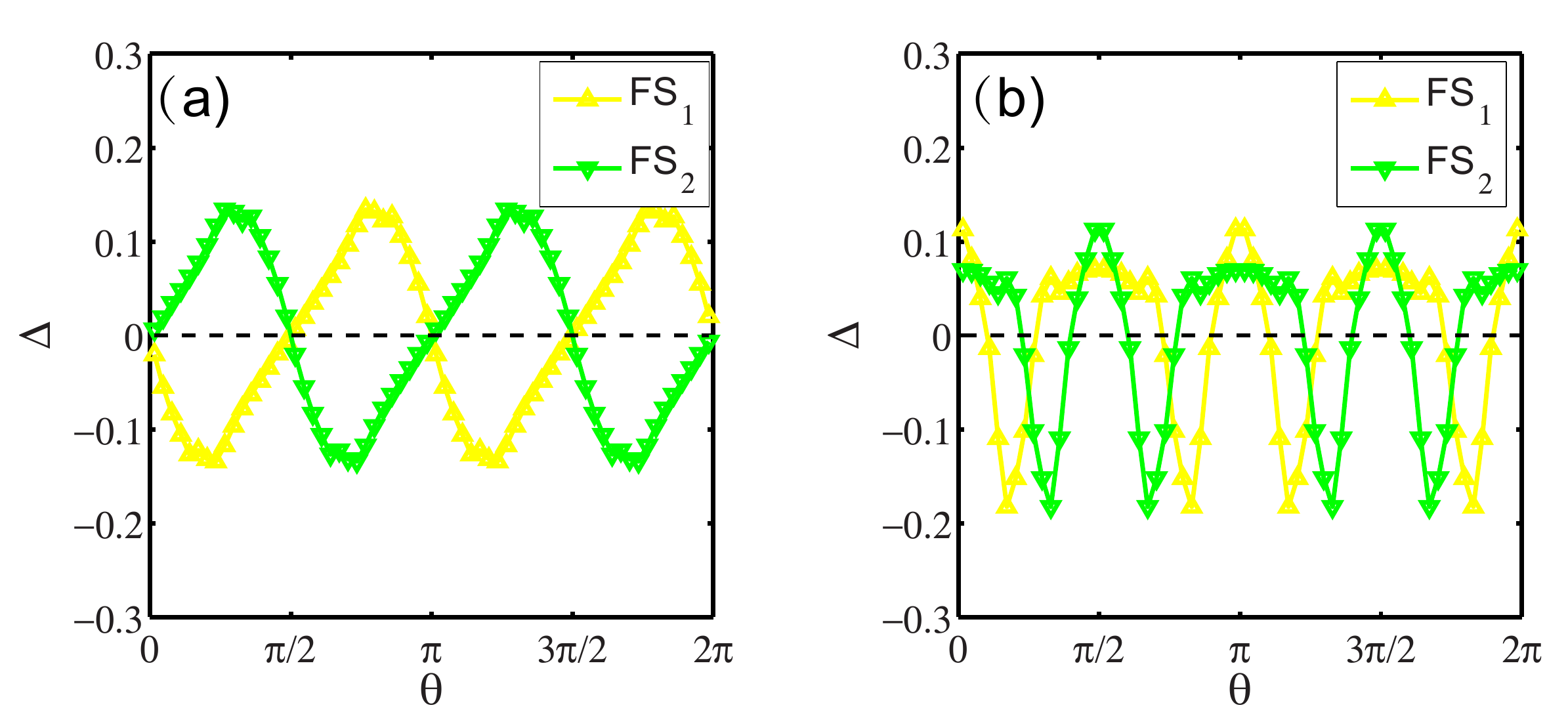}
 \caption{ (Color online) Gap form factors for the two leading pairing channels for the electron doped compound($x$=0.14) with U=2.5 and J=0.25 in FRG calculations. (a) The leading $A_{2g}$ gap. (b) The subleading $A_{1g}$ gap. $\theta$ is the polar angle on each FS with respect to its center and $\theta=0$ indicates the $+k_x$ direction. The yellow (upward triangles) and green(downward triangles) lines label the two electron FS centered at ($\pi$,0) and (0,$\pi$), respectively. }\label{frg_gap}
 \end{figure}

We use FRG calculations to justify the above RPA results.  The FRG approach is described in Ref.\cite{Wang2009,Zhai2009}. Each electron pocket is discretized into 128 patches. With U=2.5 and J=0.25, we find a very weak pairing divergence for electron doped system($x$=0.14), shown in Fig.\ref{frg_flow}. Although there is no instability in FRG flow, the calculation can still tell us that g-wave $A_{2g}$ state is more likely to be developed than s-wave $A_{1g}$ state, which is consistent with the results of RPA. The pairing form factors of the two channels are shown in Fig.\ref{frg_gap}(a) and (b), which are quite similar to the corresponding gaps obtained from RPA.  The FRG result shows that  the pairing strength of the g-wave state become weaker with the increasing of $J$ but it is always the leading one for $J<0.5$. We also perform calculation with larger U and find that the g-wave state is always the leading one if $J/U<0.2$. When $J/U>0.2$, the leading pairing symmetry becomes the $d_{x^2-y^2}$-wave ($B_{1g}$ state). The g-wave state is robust in a wide range of electron doping before the Lifshitz transition if $J$ is relative small. However, the leading pairing symmetry becomes  the s-wave $A_{1g}$ close to the Lifshitz transition point. The subleading pairing symmetry is the $d_{x^2-y^2}$-wave, which slightly differs from the one obtained by  RPA calculations.

As the Hund's coupling is relatively weak,  the g-wave is robustly favored for  BiS$_2$ superconductors at low doping. The superconducting gap is quite similar to that of g-wave state proposed for cuprate\cite{Shevchenko1998}.  In fact,  the low doping region is the only region that a single g-wave pairing  can stand out.   If one checks the pairing symmetry near Lifshitz transitions, all results show that no pairing symmetry is clearly favored\cite{Matrins2013}. At high doping region, two degenerate $A_{1g}$ and $B_{2g}$ states compete with each other in the RPA calculation. While, a similar $A_{1g}$ state is obtained but no competing $B_{2g}$ state in Ref.\cite{Liang2014}, where the strong coupling t-J model is adopted. This indicates $B_{2g}$ state is suppressed with the increasing of interaction. As our calculations mainly focus on the experimental low doping levels, the pairing symmetry is different from the previous studies that focus on relative high doping levels\cite{Usui2012,Matrins2013,Yang2013}. Actually, near the Lifshitz transition point, the obtained pairing symmetry is consistent with that of Ref.\cite{Matrins2013}.

In the above calculation, we do not consider spin orbital coupling in Bi. Due to the absence of $p_{Z}$ orbitals, the spin orbital coupling does not involve spin-flips and spin is still a good quantum number. The spin orbital coupling has little effect on the topology of FSs and the distribution of orbital characters on FSs . Therefore, we should expect that spin orbital coupling has little effect on the pairing symmetry in BiS$_2$ superconductors.

Can we find other materials with g-wave pairing symmetry? Like the high Tc materials, we consider materials containing $d$ orbitals. The correlation effect in $d$ orbitals is much stronger than that of $p$ orbital. We consider a material where the states near the Fermi level are mainly contributed by $d_{xz}$ and $d_{yz}$ orbitals due to the crystal field splitting. The Fermi surfaces and orbital characters($d_{xz}\sim p_x$, $d_{yz}\sim p_y$) are similar to those in Fig.\ref{FS_sus}(c), where electron or hole pockets are around X point. If this system becomes superconducting, the pairing symmetry may be $g$-wave. This can help us to find g-wave superconducting materials.

The g-wave pairing state can be easily justified or falsified by experiments. The most distinct feature is  the symmetry protected nodes on Fermi surfaces as shown Fig.\ref{UJ0}(b). There are 8 nodal points on Fermi Surfaces. The high resolution ARPES can directly probe the nodal structure. Physical properties related to low energy excitations, such as thermal conductivity, spin relaxation and penetration depth, should be very similar to the d-wave state in cuprates.

In summary, we have studied the pairing properties of BiS$_2$ based superconductors at low electron doping level using RPA and FRG methods. Our calculations suggest that a $g$-wave(A$_{2g}$) state is dominant at electron doping $x<0.25$ when Hund's rule coupling is relative small compared with intra-orbital Coulomb interactions($J\leq0.2U$). The g-wave state can be falsified by its distinctive nodal structures. A proof of the g-wave will not only crown BiS$_2$ as the first superconductor with the g-wave pairing, but also shed light on the mechanism of unconventional superconductors.

\acknowledgments

The work is supported by "973" program (Grant No.
 2010CB922904 and No. 2012CV821400), as well as  national science foundation of China (Grant No. NSFC-1190024, 11175248 and 11104339).


\begin{thebibliography}{bis2}
\bibitem{Mizuguchi2012} Y. Mizuguchi, H. Fujihisa, Y. Gotoh, K. Suzuki, H. Usui, K.
Kuroki, S. Demura, Y. Takano, H. Izawa and O. Miura, Phys. Rev. B {\bf 86}, 220510 (2012).

\bibitem{Singh2012} S. K. Singh, A. K., B. Gahtori, G. Sharma, S. Patnaik, and V. P. S. Awana, J. Am. Chem. Soc. {\bf 134} 16504(2012).
\bibitem{Mizuguchi2} Y. Mizuguchi, S. Demura, K. Deguchi, Y. Takano, H. Fujihisa,
Y. Gotoh, H. Izawa, and O. Miura, J. Phys. Soc. Jpn. {\bf 81}, 114725 (2012).
\bibitem{Demura} S. Demura, Y. Mizuguchi, K. Deguchi, H. Okazaki, H. Hara,
T. Watanabe, S. Denholme, M. Fujioka, T. Ozaki, H. Fujihisa, Y. Gotoh, O. Miura, T. Yamaguchi, H. Takeya, Y. Takano,
J. Phys. Soc. Jpn., {\bf 82}, 033708 (2013).
\bibitem{Xing} J. Xing, S. Li,X. Ding, H. Yang, and H.-H.Wen, Phys. Rev. B {\bf 86}, 214518 (2012).
(2012).
\bibitem{Jha} R. Jha, S. Singh, and V. P. S. Awana, J. Supercond. Novel
Magn. {\bf 26}, 499 (2013).

\bibitem{Lin2013} X. Lin, X. X. Ni, B. Chen, X. F. Xu, X. X. Yang, J. H. Dai, Y.
K. Li, X. J. Yang, Y. K. Luo, Q. Tao, G. H. Cao, and Z. A. Xu,
Phys. Rev. B {\bf 87}, 020504(R) (2013).

\bibitem{Yazici2013} D. Yazici, K. Huang, B. D. White, I. Jeon, V. W. Burnett, A.
J. Friedman, I. K. Lum, M. Nallaiyan, S. Spagna, and M. B.
Maple, Phys. Rev. B {\bf 87}, 174512 (2013).

\bibitem{Deguchi2013} K. Deguchi, Y. Mizuguchi, S. Demura, H. Hara, T. Watanabe,
S. J. Denholme, M. Fujioka, H. Okazaki, T. Ozaki, H. Takeya,
T. Yamaguchi, O. Miura, and Y. Takano, Europhys. Lett. {\bf 101}, 17004 (2013).
 \bibitem{Biswas2013}   P. K. Biswas, A. Amato, C. Baines, R. Khasanov, H. Luetkens, Hechang Lei, C. Petrovic, E. Morenzoni, arxiv:1309.7282 (2013).
\bibitem{YKLi2014} Y. K. Li, X. Lin, L. Li, N. Zhou, X. F. Xu, C. Cao, J. H. Dai, L. Zhang, Y. K. Luo, W. H. Jiao, Q. Tao, G. H. Cao, Z. Xu, Supercond. Sci. Technol. {\bf 27},035009 (2014).



\bibitem{Usui2012} H. Usui, K. Suzuki and K. Kuroki. Phys. Rev. B {\bf 86}, 220501 (2012).

\bibitem{Wan2013} X. G. Wan, H. C. Ding, Sergey Y. Savrasov and C. G. Duan, Phys. Rev. B {\bf 87},115124 (2013).

\bibitem{Li2013} B. Li, Z. W. Xing and G. Q. Huang. Europhys. Lett. {\bf 101}, 47002 (2013).

\bibitem{Yildirim2013} T. Yildirim, Phys. Rev. B {\bf 87}, 020506(R) (2013).

\bibitem{Lee2013} J. Lee, M. B. Stone, A. Huq, T. Yildirim, G. Ehlers, Y. Mizuguchi, O. Miura, Y. Takano, K. Deguchi, S. Demura, and S.-H. Lee, Phys. Rev. B {\bf 87}, 205134 (2013).


\bibitem{SLi2013} S. Li, H. Yang, D. Fang, Z. Wang, J. Tao, X. Ding, and H. H.
Wen, Sci. China-Phys. Mech. Astron. {\bf 56}, 2019 (2013).
\bibitem{Liu2013} J. Z. Liu, D. L. Fang, Z. Y. Wang, J. Xing, Z. Y. Du, X. Y. Zhu,
H. Yang, and H. H. Wen, arXiv:1310.0377.

\bibitem{Zhou2013} T. Zhou and Z. D. Wang, J. Supercond. Novel Magn. {\bf 26}, 2735(2013).
\bibitem{Liang2014} Y. Liang, X. X. Wu, W. F. Tsai and J. P. Hu. Front. Phys. {\bf 9} 194(2014).
\bibitem{Matrins2013} G. B. Martins, A. Moreo, and E. Dagotto, Phys. Rev. B {\bf 87},081102(R) (2013).
\bibitem{Yang2013} Y. Yang, W. S. Wang, Y. Y. Xiang, Z. Z. Li, and Q. H. Wang, Phys. Rev. B {\bf 88}, 094519 (2013).


\bibitem{Zeng2014} L. K. Zeng, X. B. Wang, J. Ma, P. Richard, S. M. Nie, H. M. Weng, N. L. Wang, Z. Wang, T. Qian, and H. Ding. arxiv:1402.1833.
\bibitem{Ye2014} Z. R. Ye, H. F. Yang, D. W. Shen, C. J. Jiang, X. H. Niu, D. L. Feng, Y. P. Du, X. G. Wan, J. Z. Liu, X. Y. Zhu, H. H. Wen, and M. H. Jiang. Arxiv:1402.2860.

\bibitem{Graser2009} S. Graser, T. A. Maier, P. J. Hirschfeld and D. J. Scalapino,  New J. Phys. {\bf 11} 025016(2009).
\bibitem{Kemper2010} A. F. Kemper, T. A. Maier, S. Graser, H. P. Cheng, P. J. Hirschfeld and D. J. Scalapino,  New J. Phys. {\bf 12} 073030(2010).
\bibitem{Bickers1989} N. E. Bickers, D. J. Scalapino and S. R. White, Phys. Rev. Lett. {\bf 62} 961(1989).
\bibitem{Kubo2007} K. Kubo, Phys. Rev. B {\bf 75} 224509 (2007).

\bibitem{Wang2009} F. Wang, H. Zhai, Y. Ran, A. Vishwanath, and D. H. Lee,  Phys. Rev. Lett., {\bf 102}, 047005(2009).
\bibitem{Zhai2009} H. Zhai, F. Wang and D. H. Lee, Phys. Rev. B, {\bf 80} 064517(2009).

\bibitem{Shevchenko1998} P. V. Shevchenko, and O. P. Sushkov, Physica C, {\bf 295}, 292(1998).

\end{thebibliography}
\end{document}